\documentclass[pdflatex,sn-mathphys-num]{sn-jnl}

\usepackage{graphicx}%
\usepackage{multirow}%
\usepackage{amsmath,amssymb,amsfonts}%
\usepackage{amsthm}%
\usepackage{mathrsfs}%
\usepackage[title]{appendix}%
\usepackage{xcolor}%
\usepackage{textcomp}%
\usepackage{manyfoot}%
\usepackage{booktabs}%
\usepackage{algorithm}%
\usepackage{algorithmicx}%
\usepackage{algpseudocode}%
\usepackage{listings}%
\usepackage{gensymb} 
\usepackage{hyperref}

\theoremstyle{thmstyleone}%
\theoremstyle{thmstyletwo}%

\theoremstyle{thmstylethree}%

\begin{document}

\title[Article Title]{Computational Analysis of the Temperature Profile Developed for a Hot Zone of 2500℃ in an Induction Furnace}


\author*[1]{\fnm{Juan C.} \sur{Herrera}}\email{vinod.kumar@tamuk.edu}
\equalcont{These authors contributed equally to this work.}

\author[1]{\fnm{Laura} \sur{Sandoval}}\email{llsandoval@miners.utep.edu}
\equalcont{These authors contributed equally to this work.}

\author[1]{\fnm{Piyush} \sur{Kumar}}\email{pkumar2@utep.edu}
\equalcont{These authors contributed equally to this work.}

\author[1]{\fnm{Sanjay S.} \sur{Kumar}}\email{sshanthakumar@utep.edu}
\equalcont{These authors contributed equally to this work.}

\author[1]{\fnm{Arturo} \sur{Rodriguez}}\email{arodriguez123@miners.utep.edu}
\equalcont{These authors contributed equally to this work.}

\author[2]{\fnm{Vinod} \sur{Kumar}}\email{jcherrera8@miners.utep.edu}
\equalcont{These authors contributed equally to this work.}

\author[1]{\fnm{Arturo} \sur{Bronson}}\email{abronson@utep.edu}
\equalcont{These authors contributed equally to this work.}

\affil*[1]{\orgdiv{Department of Aerospace and Mechanical Engineering}, \orgname{The University of Texas at El Paso}, \orgaddress{\street{500 W University Ave}, \city{El Paso}, \postcode{79968}, \state{Texas}, \country{United States}}}

\affil[2]{\orgdiv{Department of Mechanical and Industrial Engineering}, \orgname{Texas A\&M University at Kingsville}, \orgaddress{\street{700 University Blvd}, \city{Kingsville}, \postcode{78363}, \state{Texas}, \country{United States}}}


\abstract{Temperature gradients developed at ultra-high temperatures create a challenge for temperature measurements that are required for material processing. At ultra-high temperatures, the components of the system can react and change phases depending on their thermodynamic stability. These reactions change the system's physical properties, such as thermal conductivity and fluidity. This phenomenon complicates the extrapolation of temperature measurements, as they depend on the thermal conductivity of multiple insulating layers. The proposed model is an induction furnace employing an electromagnetic field to generate heat reaching 2500\textdegree C. A heat transfer simulation applying the finite element method determined temperatures and verified experimentally at key locations on the surface of the experimental setup within the furnace. The computed temperature profile of cylindrical graphite crucibles embedded in a larger cylindrical graphite body surrounded by zirconia grog is determined. When compared to experimental results, the simulation showed a percentage error of approximately 3.4\%, confirming its accuracy.}

\keywords{Heat Transfer, Induction Furnace, Temperature Gradients, Finite Element Method}



\maketitle

\section{Introduction}\label{sec1}

Ultra-high temperature ceramics (UHTCs) are key to the future of high-speed technologies in space exploration and hypersonic vehicles \cite{bib1, bib2}.  High demand exists for synthesizing these materials which require melting points greater than 2000\textdegree C. These materials must be able to sustain extreme environments that can cause reactivity, oxidation, and vaporization \cite{bib3}.  A challenge for material processing of UHTCs is achieving the correct ultra-high temperatures that are required for the material. An induction furnace can achieve these challenging temperatures but due to the electromagnetic field, a thermocouple cannot be used to measure the internal temperature of the system as it’s done with other furnaces. The objective of the study is to simulate the internal heat generation developed in an induction furnace as heat flowing out of the system. This helped to determine the temperature profile developed within a graphite body containing UHTC samples. This is  essential in determining the exact processing temperature of the samples. 
\\ \\
The numerical simulation software ANSYS Workbench is used to simulate these extreme environments at steady-state conditions to find different temperature profiles inside the control volume model. That are validated using experimental data. An induction furnace is modeled strictly as a heat generator for simplicity in our numerical simulation. The physical model includes various physical phenomena that were not addressed in the simulation model to simplify the model's complexity. 

\section{Materials and Methods}\label{sec2}

\subsection{Physical Experiment}\label{subsec2}

To reach processing temperatures greater than 2000\textdegree C, in approximately 1 hour, a 50kW-RDO furnace with a frequency range of 10--20kHz generates an electromagnetic field for heating. The experimental setup consists of a quartz beaker housing a graphite closed-one end tube serving as an enclosure of crucibles all within a 4-turn Cu coil. An insulative layer of zirconia grog insulates the quartz crucible from the heat generated within the graphite enclosure. The experimental components consist of an Hf-Nb-Ti alloy placed in cylindrical cavities in a graphite rod to serve as crucibles at the hot region of the furnace. An inverted graphite crucible and a liquid metal seal at the lower end of the graphite enclosure ensures a low oxygen partial pressure of less than a $p_{O_2}$ of $10^{-18}$ atm preventing oxidation of components. The pseudo-isopiestic technique used to control the oxygen potential with a temperature gradient in an induction furnace with the hot zone at 1860\textdegree C is described by Maheswaraiah, Sandate, and Bronson \cite{bib3}.
\\ \\
In the present study, a graphite spacer is placed between the UHTC samples and the liquid metal to increase the temperature gradient between the hot region of 2500\textdegree C and the cold region approximating 1200\textdegree C within the graphite enclosure. The cold regions were generated through cooling at the base and perimeter of the quartz crucible as shown in Figure~1.

\begin{figure}[h!]
    \centering
    \includegraphics[width=0.65\textwidth]{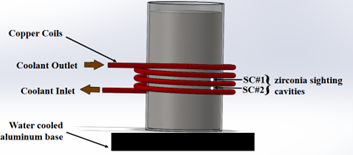} 
    \par\vspace{0.5em}
    \textbf{Fig. 1} Simplified Induction Furnace Model 
\end{figure}

\noindent While the fluctuating magnetic field polarizes metals on an ionic level creating friction and generating heat, the cooling systems cool regions of interest and create a temperature gradient throughout the system. This temperature gradient is heavily dependent on the insulative properties of materials used.  
\\ \\
\noindent For this system, the samples are contained inside of a graphite crucible. This crucible is inside another crucible which is surrounded by Zirconia (\( \text{ZrO}_2 \)) grog. The furnace temperatures are recorded using sighting cavities made of zirconia, as illustrated in Figure 1. The computer-aided design (CAD) model used for this simulation is shown in Figure 2. The model is reduced to a quarter of its size to reduce the computational cost with the assumption that it is an asymmetrical model. This is true when considering parts within the graphite crucible walls. A cross-sectional view is shown in Figure 2(c) which is further shown in detail in Figure 3. To safeguard against atmospheric air reactions, the samples are shielded by small graphite caps and an inverted graphite crucible. The model depicts the placement of two thermal-sighting cavities, a graphite crucible containing UHTC samples, Ta foil, Ti foil, an insulative layer of \( \text{ZrO}_2 \) grog, and graphite susceptors used to absorb the electromagnetic field. The UHTC sample is composed of a solid graphite cylinder with holes that serve as cavities to hold the Hf-Nb-Ti alloy situated on \( \text{B}_4\text{C} \) powder as previously shown by Bronson et al.\cite{bib5}. Temperatures readings over an operating period have been recorded and plotted to confirm the high temperature profiles. Illustrated in Figure 4.

\begin{figure}[h!]
    \centering
    \includegraphics[width=0.65\textwidth]{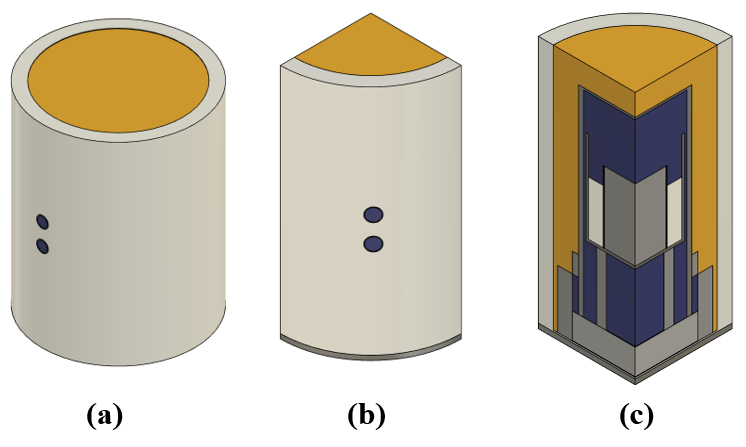} 
    \par\vspace{0.5em}
    \textbf{Fig. 2} CAD Model of Experiment (a) Isometric View (b) Isometric View of the Quarter Model (c) Cross-Sectional View 
\end{figure}

\begin{figure}[h!]
    \centering
    \includegraphics[width=0.65\textwidth]{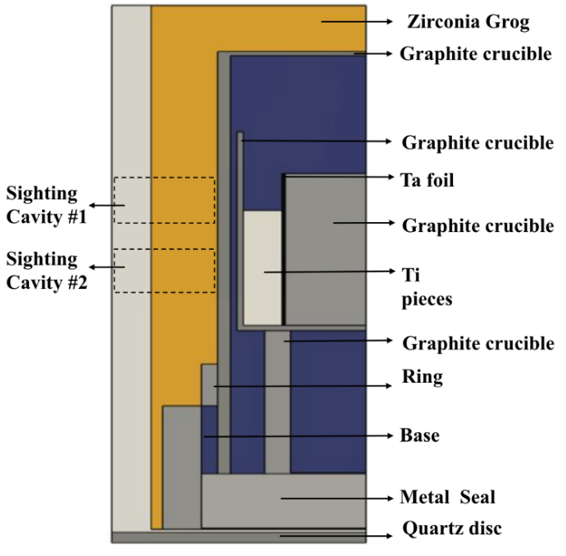} 
    \par\vspace{0.5em}
    \textbf{Fig. 3} Different Layers in the Experimental Set-Up 
\end{figure}

\begin{figure}[h!]
    \centering
    \includegraphics[width=0.65\textwidth]{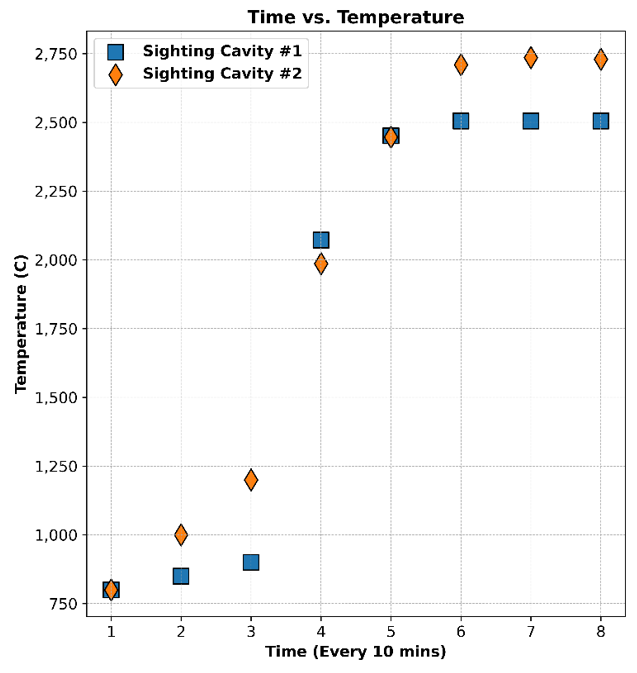} 
    \par\vspace{0.5em}
    \textbf{Fig. 4} Recorded Temperatures of the Two Sighting Cavities that were Acquired during the Experimental Run 
\end{figure}

\noindent Blackman et al. \cite{bib4} have also performed physical experiments under these conditions, where an induction furnace is also used to achieve an extreme temperature of 3000 \textdegree C in an Ar atmosphere. In their experiments, the electromagnetic field contributed to the ionization resulting in arcing which is controlled by increasing the gas pressure above one atmosphere. Maheswaraiah, Sandate, and Bronson \cite{bib3} also experienced the troublesome arching, but they minimized it by changing from an Ar to a He atmosphere and by decreasing its flow rate to 30 ml/min. 

\subsection{Numerical Simulation}\label{subsec2}

ANSYS Workbench, is used to solve the time-dependent partial differential heat equation below:

\begin{equation}
\frac{\partial u}{\partial t} = K \nabla^2 u + f
\end{equation}

\begin{equation}
u = u(t, x)
\end{equation}

\noindent Here $u$ is temperature, which is a function of space and time. The spatial Laplacian operator ($\nabla^2$) is applied to the temperature variable to describe how the temperature field varies in space. Representing the spatial distribution of temperature gradients by measuring the rate at which temperature is changing at a particular point in space. This is accompanied by an additional forcing term in the equation, which is expressed as heat generation. Since only a steady-state thermal analysis is being addressed, the transient term becomes negligible, allowing the establishment of the steady-state heat equation.
\\ \\
\noindent This strong form equation gets transformed into the weak form of the equation by integrating by parts and multiplying by the test function $v \in \hat{V}$, where the solution is valid for any test function:

\begin{equation}
a(u, v) = L_{n+1}(v)
\end{equation}

\noindent Where:

\begin{equation}
a(u, v) = \int_{\Omega} \left( uv + \Delta t K \nabla u \cdot \nabla v \right) \, dx
\end{equation}

\begin{equation}
L_{n+1}(v) = \int_{\Omega} \left( u^n + \Delta t f^{n+1} \right) \cdot v \, dx
\end{equation}

\noindent Equally, this can also be formulated via finite difference:

\begin{equation}
\frac{u^{n+1} - u^n}{\Delta t} = K \nabla^2 u^{n+1} + f^{n+1}
\end{equation}

\noindent The solution to the thermal analysis problem is obtained through an iterative process over time. Leveraging the robustness of the Finite Element Method (FEM), the ANSYS Thermal package is employed for solving the problem.  The problem at hand has been intentionally simplified to focus solely on a steady-state thermal analysis, aligning with previous studies conducted \cite{bib6, bib7}.

\subsection{Simplified Analytical Heat Conduction}\label{subsec2}

\noindent As the induction furnace generates heat through the electromagnetic conduction of metal, the hot zone is anticipated to be in the titanium-rich region. Since graphite also readily conducts the electromagnetic field, it is an additional source of heat generation. A simplified heat transfer analysis is used to  determine the processing temperature of UHTC samples and heat flow in the system. This is done by applying Fourier’s Law to the geometry shown in Figure 5, using the data in Table 1. Fourier’s law of heat conduction is used to model heat transfer through various cylindrical layers as shown in Equation (7). The application of Fourier's law of heat conduction, can be expressed as:

\begin{equation}
\dot{Q} = -kA \frac{dT}{dr} \, (W)
\end{equation}
\\
Equation (7) is separated and integrated from $r = r_1$, where $T(r_1) = T_1$, to $r = r_2$, where $T(r_2) = T_2$, to produce the following equation:

\begin{equation}
\int_{r = r_1}^{r_2} \frac{\dot{Q}}{A} \, dr = - \int_{T = T_1}^{T_2} k \, dT
\end{equation}

\noindent Substituting $A = 2\pi r L$ as the cross-sectional area and performing the integrations gives the following:

\begin{equation}
\dot{Q} = 2\pi L k \frac{T_1 - T_2}{\ln \left( \frac{r_2}{r_1} \right)} \, (W)
\end{equation}
Where:
\begin{equation}
\dot{Q} = \frac{T_1 - T_2}{R_{\text{cyl}}}
\end{equation}
Where:
\begin{equation}
R_{\text{cyl}} = \frac{\ln \left( \frac{r_2}{r_1} \right)}{2\pi L k}
\end{equation}

\noindent Equation (9) is obtained by solving the heat conduction equation of a cylindrical body with insulated layers and obtaining the temperature distribution. Temperature profiles are exterior faces of the crucibles located inside the hot zone that are expressed as $T_1 = 2500^\circ \text{C}$ and $T_6 = 2100^\circ \text{C}$ located at the exterior face of the inverted crucible shown in Figure 5(a). Representing the recorded temperature of sighting cavity 1 shown in Figure 4. Resistance is calculated for each layer of the model using Equation (11) above. This process is iterated for each layer as shown in Figure 5(a) in regions within $T_1$ and $T_6$. Each layer is considered a resistor within this thermal circuit. After calculating the total thermal resistance within the cylindrical layers, Equation (10) is used to obtain a heat transfer rate of 1319 W from the outer crucible face\cite{bib8, bib9}. These parameters are used as initial boundary conditions for the simulation.
\\ \\
\noindent A top and cross-sectional area view of the experimental set up is shown in Figure 5. This is used to conduct the heat transfer analysis described.  Each radius represents a body that is of different lengths(L1-L4) as shown in Figure 5(b). Table 1 shows the parameters that are used in the heat transfer analysis.

\begin{figure}[h!]
    \centering
    \includegraphics[width=0.65\textwidth]{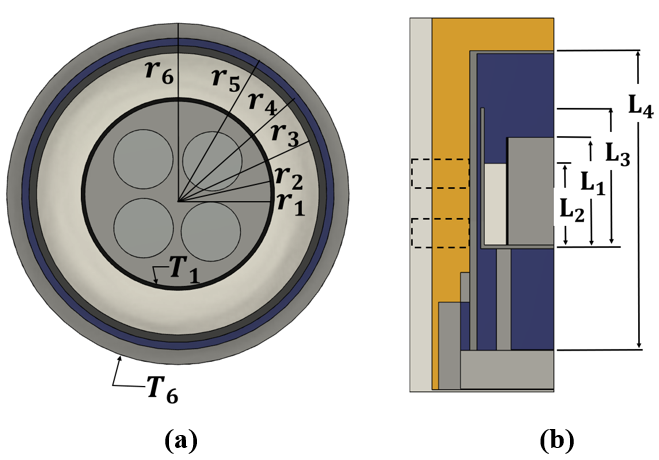} 
    \par\vspace{0.5em}
    \textbf{Fig. 5} Heat Conduction Geometry (a) Top View of the Insulated Layers (b) Cross-sectional View Showing Lengths of Cylinders Internally 
\end{figure}

\begin{figure}[h!]
    \centering
    \textbf{Table 1:} Geometry Dimensions and Thermal Properties used to Calculate Heat Flow 
    \par\vspace{0.5em}
    \includegraphics[width=0.65\textwidth]{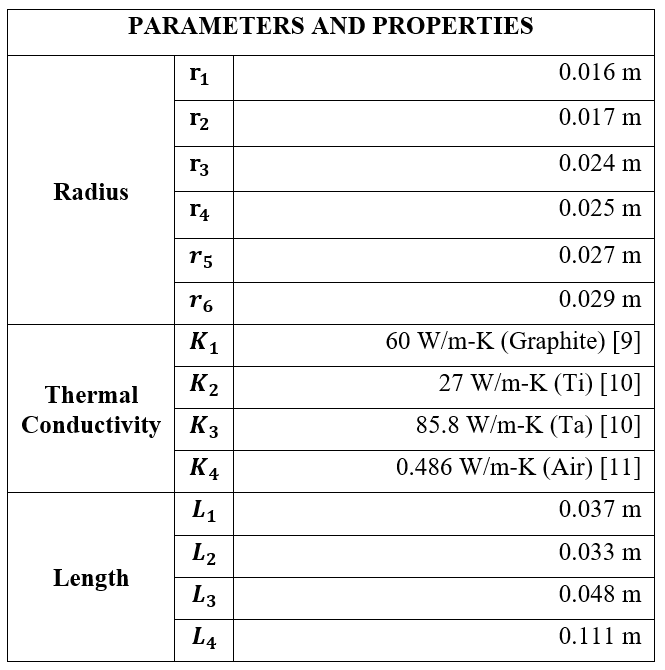} 
\end{figure}

\noindent The different thermal conductivity values for the materials used in the heat transfer analysis are evaluated at high temperatures and can be shown in Table 1. Thermal conductivity values for Ta \cite{bib10} and Air \cite{bib11} were evaluated at 2500\textdegree C. Thermal conductivity values that were not available at the specific temperature of 2500 \textdegree C were evaluated at their highest temperature of either 1800\textdegree C and 2000\textdegree C, which consist of Ti \cite{bib10} and Graphite \cite{bib9, bib12}.

\begin{figure}[h!]
    \centering
    \includegraphics[width=0.72\textwidth]{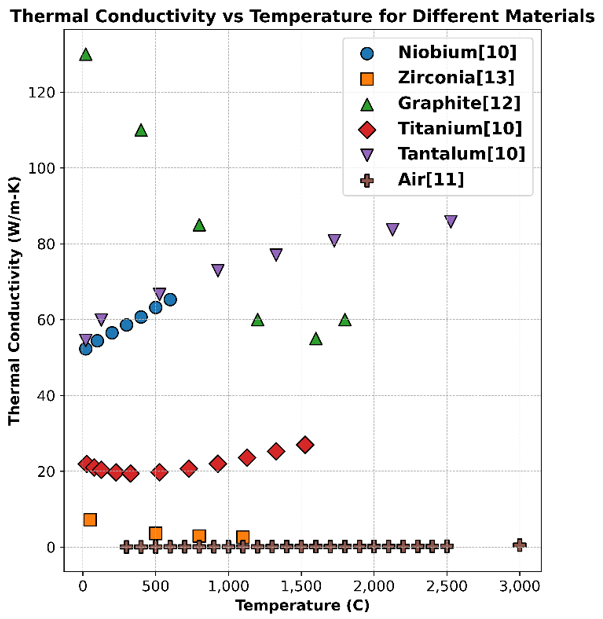} 
    \par\vspace{0.5em}
    \textbf{Fig. 6} Thermal Conductivity of Materials vs. Temperature 
\end{figure}

\noindent Thermal conductivity values at high temperatures that are used in the simulation are shown in Figure 6, which include materials Niobium \cite{bib10}, and Zirconia \cite{bib13}.

\section{Results}\label{sec3}

\subsection{Initial Boundary Conditions for Simulation}\label{subsec3}

The sighting cavities, denoted as A and B in Figure 7, represent SC\#1 and SC\#2, respectively. These sighting cavities were strategically positioned adjacent to the anticipated hot zone for temperature measurements using pyrometers in the experimental run previously mentioned. \(T_{\text{SC\#1}} > T_{\text{SC\#2}}\) since it is placed closer to the hot zone. A third pyrometer is used to gather quartz temperature data; this point is labeled C in Figure 7. The final parameters, D and E, were calculated values from the heat transfer analysis that consist of heat flow and crucible temperature, respectively.

\begin{figure}[h!]
    \centering
    \includegraphics[width=0.70\textwidth]{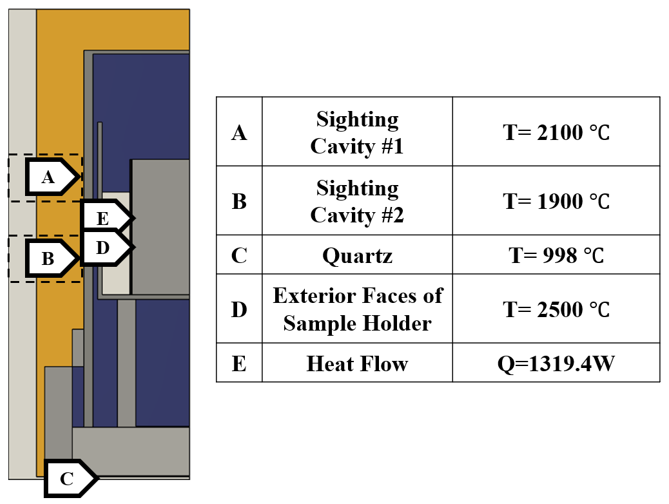} 
    \par\vspace{0.5em}
    \textbf{Fig. 7} Initial Boundary Conditions Gathered from Experimental Data 
\end{figure}

\begin{figure}[h!]
    \centering
    \includegraphics[width=0.85\textwidth]{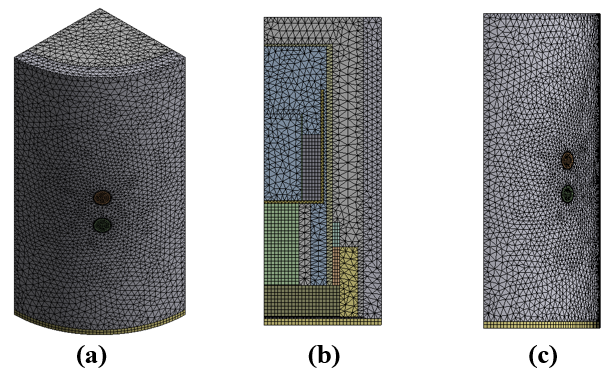} 
    \par\vspace{0.5em}
    \textbf{Fig. 8} Mesh of Model (a) Isometric View (b) Cross-Sectional Area (c) Side View 
\end{figure}

\begin{figure}[h!]
    \centering
    \textbf{Table 2:} Mesh Characteristics 
    \par\vspace{0.5em}
    \includegraphics[width=0.4\textwidth]{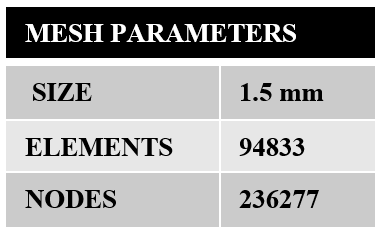} 
\end{figure}

\noindent An isometric and cross-sectional view of the quarter size model is shown in Figure 8. The mesh parameters are described in Table 2.

\subsection{Simulation Results}\label{subsec3}

\begin{figure}[h!]
    \centering
    \includegraphics[width=0.7\textwidth]{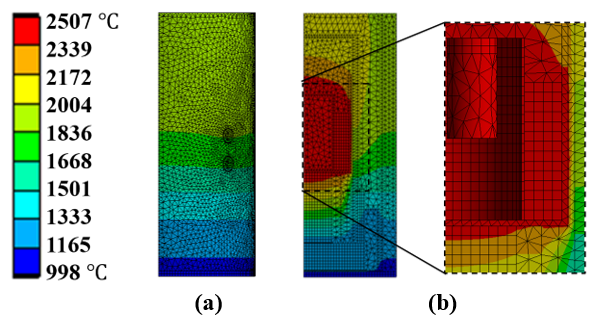} 
    \par\vspace{0.5em}
    \textbf{Fig. 9} Simulation Results showing Temperature Profiles (a) Cross-Sectional View and (b) Side View 
\end{figure}

\noindent Simulation results reveal two distinct temperature gradients as shown in the outer section of the quartz beaker, i.e., Figure 9 (a). The cooler region at the bottom of the experimental set-up draws heat toward the bottom though a temperature of 998°C is still attained.  The second gradient shown in Figure 9 (b) highlights the hot zone in red. This is identified as the hot zone due to the internal heat generation that comes from the reacting components.

\begin{figure}[h!]
    \centering
    \includegraphics[width=0.6\textwidth]{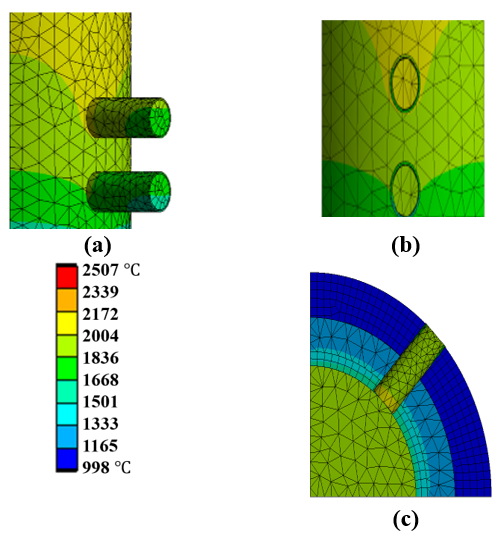} 
    \par\vspace{0.5em}
    \textbf{Fig. 10} Simulation Results showing Temperature Profiles of Sighting Cavities (a) Side View with Air and (b) Front View without Air and a Top View of the Sighting Cavities (c) 
\end{figure}

\noindent Initial conditions are established but the simulation proves there is a temperature difference between the established initial temperature profiles and the simulation results. Figure 10 focuses on the sighting cavities front and side view.
\begin{enumerate}[(a)]
\item Figure 10 shows the cylindrically modeled air which accounts for the heat transfer properties of air present in the simulation.
\item Figure 10 removes the cylindrical bodies of air for an unobstructed view of the sighting cavity temperature. Temperature variations are observed between the initial boundary conditions and the simulation results at the same location. From Figure 7, the sighting cavity temperatures are 2100\degree C and 1900\degree C. However, the simulation temperatures for SC\#1 and SC\#2 were 2172\degree C and 1836\degree C, respectively. These values deviated as follows:
\[
\Delta T_{\text{SC\#1}} = 72\degree C, \quad \Delta T_{\text{SC\#2}} = 64\degree C.
\]
The values for both SC\#1 and SC\#2 exhibit an error of approximately 3.4\%.
\end{enumerate}

\section{Discussion}\label{sec4}

A simplified heat transfer analysis is used to determine the heat flow and internal temperature using Equation 10, which is used for comparison with experimental values obtained from the sighting cavities. Combined, these values served as the initial boundary conditions for a steady-state simulation illustrated in Figure 9. The simulation showed a vertical temperature gradient due to the water-cooled base, with a temperature drop of approximately 1200\degree C as depicted in Figure 9. This gradient verified the behavior of the hot zone in relation to the grog insulation layers, with most of the heat concentrated in the central region. 
\\ \\
Notably, the sighting cavities aimed to capture temperature profiles under the assumption that recorded temperatures mirrored those within the hot zone. However, a temperature disparity emerged post-simulation of 
\[
\Delta T_{\text{SC\#1}} = 72\degree C, \quad \Delta T_{\text{SC\#2}} = 64\degree C
\]
for SC\#1 and SC\#2, respectively. This is likely due to disagreements in the reported thermal conductivities at temperatures above 2000\degree C. Thermal properties at these temperatures are scarce and vary significantly between authors. Additionally, at elevated temperatures, materials experience phase changes which directly affect the thermal properties of the system. 
\\ \\
This error is minimized by placing sighting cavities close to the UHTC sample. The SC placement allows direct temperature measurements of the outer graphite crucible. This reduces errors such as those caused by impurities in the ZrO\textsubscript{2} grog and phase changes in quartz.
\\ \\
Temperature readings for the error calculations were acquired via temperature probes within the ANSYS simulation. These readings were taken from each face of the cylindrical air model facing the ambient environment. These temperatures were approximately 800\textdegree C hotter than the surrounding zirconia grog due to the lapse in insulative material required to collect temperature measurements shown in Figure 10 (c). 
\\ \\
Air is modeled as a solid cylindrical entity inside both sighting cavities, temperature profiles for this are as shown in Figure 10. The anticipated temperature gradient between sighting cavities A and B is 100\textdegree C yet the simulation shows a gradient of 168\textdegree C (Figure 9). The heat flow and conductivity of the air is the reason the proposed temperature is slightly higher.
\\ \\
Figure 9 displayed a marginal temperature increase in the hot zone due to the heat flow rate, suggesting that the hot zone's actual temperature may differ from the assumed 2500\textdegree C when the sighting cavity records 2100\textdegree C. While direct temperature measurement in the hot zone are unavailable, transitional metals with melting points of 2233\textdegree C and 2477\textdegree C were successfully melted. This confirmed that temperatures above 2000\textdegree C were reached.

\section{Conclusion}\label{sec5}

The dynamic nature of heat transfer and material-specific thermal properties introduce a complexity in temperature recordings, requiring recalibration of initial assumptions. The simulation results show an approximate error of 3.4\% at both SC\#1 and SC\#2. The temperature profiles revealed a temperature difference of 72\textdegree C and 64\textdegree C in each sighting cavity. The hot zone’s behavior is verified by the vertical temperature gradient of 1200℃ created by the insulated layers. Despite using physical temperature recordings as established uniform temperature in the steady-state simulation. Simplified heat transfer analysis is conducted to simulate the heat generated by the induction furnace as heat flow created from the metals located in the hot zone. The observed non-uniform distribution of thermal energy influenced by the calculated heat flow rate from the hot zone created distinct temperature gradients in the model. This study contributes to the discourse of heat transfer in complex environments for material processing, emphasizing the importance of refining modeling approaches. Highlighting the understanding of thermodynamics and thermal conductivity of materials used in high temperatures in similar experimental setups. 

\section*{Acknowledgements}
We acknowledge the U.S. Department of Defense (AFOSR Grant Number \# FA9550-22-1-0018, FA9550-19-1-0304, FA9550-17-1-0393 FA9550-12-1-0242 SFFP, AFTC, HAFB/HSTT, AFRL, HPCMP), U.S. Department of Energy (GRANT13584020, DE-SC0022957, DE-FE0026220, DE-FE0002407, NETL, Sandia, ORNL, NREL), Systems Plus, and several other individuals at these agencies for partially supporting our research financially or through mentorship. We would also like to thank NSF ((HRD-1139929, XSEDE Award Number ACI-1053575), TACC, DOE, DOD, HPCMP, University of Texas STAR program, UTEP (Research Cloud, Department of Mechanical Engineering, Graduate School \& College of Engineering) for generously providing financial support or computational resources. Without their generous support, completing the milestones would have been almost impossible.

\end{document}